\documentclass [12pt]{article}

\begin{document}

\vspace*{0.6cm}

\begin{center}
{\Large\bf Passive Quantum Erasure for Neutral 
Kaons\footnote{Published in {\em Symmetries in Gravity and Field Theory},
ed.~by V.~Aldaya and J.~M.~Cervero,
Ediciones Universidad de Salamanca, Salamanca, 2004, p.~223.
Contribution to the Festschrift in honor of
Prof.~Jos\'e Adolfo de Azc\'arraga for his 60th birthday,
Salamanca (Spain), June 9--11, 2003.}}\\
\vspace*{0.6cm}
A.~Bramon$^{1}$, G.~Garbarino$^{2}$ and
B.~C.~Hiesmayr$^{3}$\\
\vspace*{0.3cm}
{\footnotesize\it $^1$Grup de F{\'\i}sica Te\`orica,
Universitat Aut\`onoma de Barcelona,
E--08193 Bellaterra, Spain \\
$^{2}$Dipartimento di Fisica Teorica, Universit\`a di Torino
and INFN, Sezione di Torino\\ I--10125 Torino, Italy \\
$^{3}$Institute for Theoretical Physics, University of Vienna,
A--1090 Vienna, Austria}
\end{center}
\vspace*{0.6cm}

\begin{abstract}
Quantum marking and quantum erasure are discussed for the neutral kaon system.
Contrary to other two--level systems, strangeness and lifetime of a neutral kaon
state can be alternatively measured via an ``\emph{active}" or a ``\emph{passive}"
procedure. This offers new quantum erasure possibilities. In particular,
the operation of a quantum eraser in the ``\emph{delayed choice}" mode is clearly
illustrated.
\end{abstract}

\section{Introduction}

Since the foundation of quantum mechanics, physicists know that
the heart of Bohr's complementarity principle lies in the role
played by measurement devices. In 1982 Scully and Dr\"uhl
\cite{scully82,scully91} proposed a gedanken experiment
---since then known as the \emph{quantum eraser}--- to discuss some of
the subtle aspects of quantum measurement which are related to
still missing keys to explain the appearance of a classical world
in quantum theory. It is similarly well known that all two--level
quantum systems are in many aspects fully equivalent and admit a
unified treatment in terms of Pauli matrices, \emph{qubit} states
and SU(2) formalism. Best known examples of such two--level
systems are offered by spin--$1/2$ states, polarized photon states and
neutral kaon states. Experimental tests of quantum erasure have
been performed using (mostly, polarized) photon states
\cite{zeilinger95,walborn,scully00}. The general purpose of the
present contribution is to extent these quantum eraser
considerations to the neutral kaon system.

The basic idea behind quantum eraser experiments is that two
indistinguishable and thus interfering amplitudes of a quantum
system, the \emph{object}, can become distinguishable (``marked'')
thanks to the \emph{entanglement} with a second quantum system,
the \emph{meter}. In two--path interferometric devices, the latter
system is frequently  called  a ``which way'' detector. If the
information stored in the meter system ---a kind of quantum
\emph{mark}--- is even in principle accessible, the object system
looses all interference abilities. However, if one somehow manages
to ``erase'' this distinguishability, by correlating the outcomes
of the measurements on the object system with those of specific
measurements on the meter system, one can recover the original
object interferences.

The case in which the meter is a system distinct and spatially
separated from the object is of particular interest. Indeed, in
this case the decision to erase or not the meter ``mark'' and thus
the distinguishability of
the object amplitudes (i.e., to observe or not interference) can be
taken long after the object has been measured. Quantum erasure is
then performed in the \emph{delayed choice} mode which best captures the
essence of the phenomenon \cite{scully99,mohrhoff}.
On the other hand, the neutral kaon system shows us that the erasure operation
can be carried out actively, i.e., by exerting the free will of the experimenter,
or passively, i.e., randomly exploiting a particular
quantum--mechanical property of the meter system.

The specific purposes of this contribution are to discuss how one can
distinguish between ``\emph{active}'' and ``\emph{passive}'' quantum erasure
for neutral kaons and how one can operate in the ``\emph{delayed choice}''
mode. Entangled pairs of kaons turn out to be maximally
appropriate for these two purposes.

\section{Strangeness and lifetime measurements}\label{measurments}
\label{measurement}

Contrary to what happens with spin--$1/2$ particles or photons, neutral kaons only
exhibit two different measurement bases \cite{bg1,ABN}: the strangeness and the
lifetime bases.

The strangeness basis, $\lbrace K^0,\bar K^0\rbrace$ with $\langle K^0|\bar
K^0\rangle=0$, is the appropriate one to discuss strong production and reactions of
kaons. If a dense piece of nucleonic matter is inserted along a
neutral kaon beam, the incoming state is projected either into a $K^0$ by the
strangeness conserving strong interaction $K^0 p\rightarrow K^+ n$ or into
a $\bar K^0$ via $\bar K^0 p\rightarrow \Lambda \pi^+, \bar K^0 n\rightarrow K^-
p$ or $\bar K^0 n\rightarrow \Lambda \pi^0$. These strangeness detections are
totally analogous to the projective von Neumann measurements of a two--channel
analyzer, e.g., of polarized photons. By inserting the piece of matter along the
beam, one induces an ``\emph{active}''  measurement of strangeness.

The strangeness content of  neutral kaon states
can alternatively be detected by observing their semileptonic decay modes.
Indeed, these semileptonic decays obey the well tested 
$\Delta S=\Delta Q$ rule which allows the modes
\begin{equation}
K^0(\bar s d) \to \pi^-(\bar u d)\;+\;l^+\;+\;\nu_l \; , \; \;
\bar{K}^0(s\bar d) \to\pi^+(u \bar d)\;+\;l^-\;+\;\bar\nu_l ,
\end{equation}
where $l$ stands for $e$ or $\mu$, but forbids decays into the
respective charge conjugated modes. Obviously, the experimenter
cannot induce a kaon to decay semileptonically or even at a given
time: he or she can only sort at the end of the day all observed
events in proper decay modes and time intervals. We thus have a
``\emph{passive}'' procedure for strangeness measurements.

The \emph{active} measurement is then monitored by strangeness
conservation while the \emph{passive}  measurement is assured
by the $\Delta S=\Delta Q$ rule. The probabilities for single
kaons measured using both procedures have been proved to agree
with the quantum--mechanical predictions
\cite{PDG,CPLEARreview,BGH2}.

The second basis, the lifetime basis $\{K_S,K_L\}$, consists of
the short-- and long--lived states having well defined masses
$m_{S(L)}$ and decay widths $\Gamma_{(S)L}$. It is the appropriate
basis to discuss their propagation in free space, where
\begin{eqnarray}\label{timeevolution}
|K_{S}(\tau)\rangle = e^{-i\lambda_{S}\tau} |K_{S}\rangle ,\;\;
|K_{L}(\tau)\rangle = e^{-i\lambda_{L}\tau} |K_{L}\rangle ,
\end{eqnarray}
with $\lambda_{S(L)}=m_{S(L)}-i\Gamma_{S(L)}/2$. These states preserve their own
identity in time, but, since $\Gamma_S\simeq 579\, \Gamma_L$, the $K_S$
component extincts much faster than the $K_L$ component. 
To observe if a kaon propagates as a
$K_S$ or $K_L$ at (proper) time $\tau$, one has to detect
at which time it subsequently decays. Kaons which are observed to decay before
$\simeq \tau + 4.8\, \tau_S$ have to be identified as $K_S$'s, while those surviving
after this time interval have to be
identified as $K_L$'s. Misidentifications then reduce only to a few parts in
$10^{-3}$  \cite{bg1,eberhard,BGH1}.
Such a procedure, which necessarily has to allow for free--space propagation,
represents  an ``\emph{active}'' measurement of lifetime.

Since 40 years one knows that the neutral kaon system violates the $CP$
symmetry. 
Among other things, this implies that
the weak interaction eigenstates are not strictly orthogonal,
$\langle K_S|K_L\rangle= 2 ({\rm Re}\,\varepsilon)/(1+|\varepsilon|^2)\simeq
3.2 \cdot10^{-3}$ \cite{PDG,kabir}.
However, by neglecting these small $CP$ violation effects we can discriminate
between $K_S$'s and $K_L$'s by
leaving the kaons to propagate in  free space
and observing their nonleptonic $K_S \to 2\pi$ or $K_L \to 3\pi$ decay modes. 
This represents a ``\emph{passive}'' measurement of lifetime,
since the type of kaon decay mode ---nonleptonic in the present case,
instead of semileptonic as before--- cannot
be in any way influenced by the experimenter.

The \emph{active} and \emph{passive} lifetime measurements are efficient thanks to the
smallness of $\Gamma_L/\Gamma_S$ and $\varepsilon$,
respectively. Since $\Gamma_L/\Gamma_S\simeq |\varepsilon|\simeq {\cal O}(10^{-3})$,
both effects are very small and can be safely neglected
in our discussion.

Summarizing, we have two different experimental procedures
---\emph{active} and \emph{passive}--- to measure each one
of the only two neutral kaon observables: strangeness or lifetime.
The existence of these two alternative measurement procedures has
no analog in any other two--level quantum system. In this sense,
kaons offer new possibilities for  quantum erasure experiments,
though other non--kaonic two--level systems clearly offer more (in
principle, infinitive many) measurement bases.

\section{\emph{Active} and 
\emph{passive} joint measurements}
\label{probabilities}

We now introduce an entangled two--kaon state which is analogous to
the standard and widely used entangled two--photon states produced
via spontaneous parametric down conversion (SPDC).
Through the decay of the $\Phi(1020)$--meson resonance
\cite{handbook} or $S$--wave proton--antiproton annihilation \cite{CPLEAR}
one obtains the anti--symmetric and maximally entangled state
at time $\tau=0$:
\begin{equation}
\label{entangled}
|\phi(0)\rangle  =  \frac{1}{\sqrt 2}\left[
|K^0\rangle_l |\bar{K}^0\rangle_r - |\bar{K}^0\rangle_l |K^0\rangle_r\right]
 =  \frac{1}{\sqrt 2}\left[
|K_L\rangle_l |K_S\rangle_r - |K_S\rangle_l |K_L\rangle_r\right] ,
\end{equation}
where $l$ and $r$ denote the ``left'' and ``right'' directions of
motion of the two separating kaons. The state has been written in
the two observable bases and $CP$--violating effects are neglected.

After production, the left (right)
moving kaon evolves according to Eq.~(\ref{timeevolution}) up to time $\tau_l$
($\tau_r$) to produce the state:
\begin{equation}
\label{notnorm}
|\phi(\tau_l,\tau_r)\rangle = \frac{1}{\sqrt 2}\left\{
e^{-i(\lambda_L \tau_l+\lambda_S \tau_r)}|K_L\rangle_l|K_S\rangle_r
-e^{-i(\lambda_S \tau_l+ \lambda_L \tau_r)}|K_S\rangle_l|K_L\rangle_r\right\} .
\end{equation}
By normalizing to kaon pairs with both members surviving up to $(\tau_l,\tau_r)$,
one obtains the state:
\begin{equation}
\label{time}
|\phi(\Delta\tau)\rangle= \frac{1}{\sqrt {1+e^{\Delta\Gamma
\Delta\tau}}}\biggl\lbrace
|K_L\rangle_l|K_S\rangle_r
-e^{i \Delta m \Delta\tau} e^{{1 \over 2} \Delta \Gamma \Delta\tau}
|K_S\rangle_l|K_L\rangle_r\biggr\rbrace 
\end{equation}
($\Delta\tau=\tau_l-\tau_r$ and $\Delta\Gamma=\Gamma_L-\Gamma_S$)
or, in the strangeness basis:
\begin{eqnarray}
\label{timestrangeness}
|\phi(\Delta\tau)\rangle &=& \frac{1}{2 \sqrt {1+e^{\Delta\Gamma \Delta\tau}}}
\left\{(1-e^{i \Delta m \Delta\tau} e^{{1 \over 2} \Delta \Gamma \Delta\tau})
\left[|K^0\rangle_l|K^0\rangle_r-|\bar K^0\rangle_l|\bar K^0\rangle_r\right] \right.
\nonumber \\
&& \left. +(1+e^{i \Delta m \Delta\tau} e^{{1 \over 2} \Delta \Gamma \Delta\tau})
\left[|K^0\rangle_l|\bar K^0\rangle_r-|\bar
K^0\rangle_l|K^0\rangle_r\right] \right\} .
\end{eqnarray}
With this normalization, we work with bipartite two--level
systems as for spin--$1/2$ entangled pairs.
The analogy between state (\ref{time}) and the
polarization--entangled two--photon state $|\Psi \rangle = \left[
|V\rangle_i |H\rangle_s - e^{i \Delta \phi}|H\rangle_i |V\rangle_s
\right]/\sqrt{2}$, where $\Delta \phi$ is a relative phase under control by the
experimenter \cite{zeilinger95},  is obvious. For an
accurate description of the time evolution of entangled neutral kaon 
pairs,  see Refs.~\cite{GGW,BH1}.

In the remainder of this section we discuss the derivation of the
observable joint probabilities corresponding to \emph{active}
and \emph{passive} measurements.

\subsection{\emph{Active} measurements on both kaons}

Considering an \emph{active} strangeness measurement on both sides
means to act with the projectors
$P_i^l P_j^r$ ($P_i=|K^0\rangle\langle K^0|,|\bar
K^0\rangle\langle \bar K^0|$) onto state
(\ref{timestrangeness}). The probabilities to observe on both
sides like-- or unlike--strangeness events are: 
\begin{eqnarray}
\label{lSprob}
P\left[K^0(\tau_l),K^0(\tau_r)\right]=P\left[\bar{K}^0(\tau_l),\bar{K}^0(\tau_r)\right]
= \frac{1}{4}\big\lbrace 1-{\cal V}(\Delta\tau) \cos(\Delta m\Delta \tau)\big\rbrace , && \\
\label{uSprob}
P\left[K^0(\tau_l),\bar{K}^0(\tau_r)\right]=P\left[\bar{K}^0(\tau_l),K^0(\tau_r)\right]
= \frac{1}{4}\big\lbrace 1+{\cal V}(\Delta\tau) \cos(\Delta m\Delta
\tau)\big\rbrace , &&
\end{eqnarray}
where:
\begin{equation}
\label{visibility}
{\cal V}(\Delta\tau)=\frac{1}{\cosh(\Delta\Gamma\Delta\tau/2)}
\end{equation}
is the visibility of the $K^0$--$\bar{K}^0$ oscillations. First, we
note that for $\Delta\tau=0$ we have perfect EPR--correlations:
the like--strangeness probabilities vanish and the
unlike--strangeness probabilities take the maximum value
[${\cal V}(0)=1$]. Second,
$\Delta m\Delta\tau$ plays the same role as the relative orientation
of polarization analyzers in the entangled photon case. The
kaon mass difference $\Delta m$ introduces automatically a time
dependent relative phase between the two kaon amplitudes. However,
opposite to the photon case, the visibility decreases as
$|\Delta\tau|\rightarrow \infty$.

If one wants to measure, \emph{actively}, strangeness on the left and lifetime
on the right, one has to remove the piece of matter on the right to
allow for free kaon propagation in space.
One can then measure the following non--oscillating joint probabilities:
\begin{eqnarray}
\label{probS}
&&P\left[K^0(\tau_l),K_S(\tau_r)\right]=P\left[\bar{K}^0(\tau_l),K_S(\tau_r)\right]
=\frac{1}{2\left(1+e^{\Delta\Gamma\Delta\tau}\right)} , \\
\label{probL}
&&P\left[K^0(\tau_l),K_L(\tau_r)\right]=P\left[\bar{K}^0(\tau_l),K_L(\tau_r)\right]
=\frac{1}{2\left(1+e^{-\Delta\Gamma\Delta\tau}\right)} .
\end{eqnarray}

\subsection{\emph{Passive} measurements on both kaons}

In this case  of \emph{passive} measurements
along both beams, one allows the entangled kaon pairs
to propagate freely in space and identifies the kaon decay \emph{times} and
\emph{modes}. As discussed in detail in Ref.~\cite{ABN}, one has to measure the
joint decay rate $\Gamma(f_l,\tau_l; f_r,\tau_r)$ which is defined as
the number of left--side decays into the mode $f_l$ between $\tau_l$ and
$\tau_l + d\tau_l$ accompanied by right--side decays into the mode $f_r$
between $\tau_r$ and $\tau_r + d\tau_r$ divided by $d\tau_l$, $d\tau_r$ and
the total number of initial kaon pairs. The quantum--mechanical expressions
for this joint decay rate can be easily deduced from Eq.~(\ref{notnorm}).
Additionally, one has to compute
the four partial decay widths $\Gamma(K_{f_i}\to f_i)$, where $K_{f_i}\to
f_i$ stands for the four  identifying decay modes
$K_S \to \pi\pi$, $K_L \to \pi\pi\pi$, $ K^0 \to \pi^- l^+ \nu_l$ or
$\bar{K}^0 \to \pi^+ l^- \bar{\nu}_l$. Finally, one has to take into account the
extinction of the beams via the normalization factor
$N(\tau_l, \tau_r) \equiv e^{-(\Gamma_L +\Gamma_S)(\tau_l +\tau_r)/2}
\cosh\left[(\Gamma_L -\Gamma_S)(\tau_l -\tau_r)/2\right]$
which depends on both
$\tau_l$ and $\tau_r$.
This allows one to define all the relevant joint detection probabilities
through the relation \cite{ABN}:
\begin{equation}
\label{obs1}
P\left[K_{f_l}(\tau_l),K_{f_r}(\tau_r)\right]=\frac{\Gamma(f_l,\tau_l;
f_r,\tau_r)}{N(\tau_l, \tau_r) \Gamma(K_{f_l}\to f_l)\Gamma(K_{f_r}\to f_r)} ,
\end{equation}
where $K_{f_l}$, $K_{f_r}=K^0$, $\bar{K}^0$, $K_L$ or $K_S$.

It is easy to see that the physical
meaning and the quantum mechanical expression for the
probabilities in Eq.~(\ref{obs1}) coincide with the previously
considered probabilities in Eqs.~(\ref{lSprob})--(\ref{probL}).
However, while the latter are measured either by actively inserting or
removing a piece of nucleonic matter in the two beams, the
measurement method via Eq.~(\ref{obs1}) is completely different:
the quantum--mechanical probabilities alone decide if each one of
the two kaons of a given pair is going to be measured in the
strangeness or lifetime basis. The experimenter remains passive in
such measurements.

\subsection{\emph{Active} and \emph{passive} measurements}

One can similarly  combine an \emph{active} measurement on one side with
a \emph{passive} measurements on the other.
With the information given in the two previous subsections, it is quite easy to
reproduce, as expected, the quantum--mechanical
results of Eqs.~(\ref{lSprob})--(\ref{probL}).

\section{Quantum eraser experiments for kaons}\label{experiment}
Several quantum eraser experiments have already been performed
\cite{zeilinger95,walborn,scully00}.  They use
SPDC to produce a two--photon maximally entangled state
which is the analog of the kaon state of Eq.~(\ref{time}). One photon of the
pair is considered as the \emph{object} system. On this photon one wants to obtain
(or not) ``which way'' information ($WW$) by a suitable measurement on the \emph{meter} photon.
Different strategies are used for marking and erasing this $WW$ information. All
these experiments need a kind of double--slit mechanism in order
to allow for a ``wave like'' behaviour of the meter--object system, which then
leads to a state similar to the one of Eq.~(\ref{time}) but with $\Gamma_S=\Gamma_L=0$.
Photon stability is certainly an advantage. However, in
the neutral kaon system there is no need of such a double--slit mechanism: it is
\emph{automatically} offered by the special time
evolution of kaons.

To understand this better, let us discuss the time evolution of a single neutral kaon,
e.g.~a $|K^0(\tau)\rangle$. Just after its production it is an equal superposition of the
lifetime eigenstates,
$|K^0(\tau=0)\rangle=\lbrace |K_S\rangle+|K_L\rangle\rbrace/\sqrt{2}$,
and according to Eq.~(\ref{timeevolution}) it starts propagating in free space in the
coherent superposition:
\begin{eqnarray}
|K^0(\tau)\rangle=\frac{1}{\sqrt{2}}\lbrace
e^{-i\lambda_S \tau}|K_S\rangle+e^{-i\lambda_L \tau}|K_L\rangle\rbrace\;.
\end{eqnarray}
The $K^0$ proceeds through a single spatial trajectory comprising \emph{automatically}
(i.e., with no need of any double--slit like apparatus) the two differently propagating
components $K_S$ and $K_L$. At $\tau=0$ there is no information on which component
is propagating but for kaons surviving after some time $\tau$,  $K_S$
propagation is less likely than $K_L$ propagation. 
For kaons, this allows one to obtain ``which
width'' information ($W\mathcal{W}$) in the very same way as one can obtain
``which way'' information ($WW$) for photons passing through a double--slit
device.

Being the two--kaon states of Eq.~(\ref{time}) or (\ref{timestrangeness})
automatically given by Nature, one can play the game of quantum marking 
and eraser experiments. 
Four possible experiments are discussed in the following \cite{BGH3}. In the first
three, (a),(b) and (c), the left moving kaon is the object; on this kaon one performs
\emph{active} strangeness measurements ---placing a piece of matter--- at different
$\tau_l$--values to scan for possible $K^0$--$\bar{K}^0$ oscillations. 
The right moving kaon is the meter;
it carries $W\mathcal{W}$ information which can be actively or passively erased
by a suitable \emph{active} or \emph{passive} measurement at a fixed time
$\tau_r^0$.
In the fourth experiment (d), \emph{passive} measurements are performed on
both sides. It is not clear which kaon, the left or the right moving one, is
playing the role of the meter: this examplifies the central point of the
delayed choice erasure.

\subsection*{(a) Active eraser with \emph{active} measurements}
In a first set--up we insert \emph{active} strangeness detectors
along both beams and consider only kaon pairs which
survive up to both detectors. We clearly observe $K_S$--$K_L$ interference
in the coincident counts of the object--meter system with the visibility
${\cal V}(\tau_l-\tau_r^0)$ of Eq.~(\ref{visibility}).
More precisely, we observe fringes
for unlike--strangeness joint detections,
Eq.~(\ref{uSprob}), and
anti--fringes for like--strangeness joint detections,
Eq.~(\ref{lSprob}).
In a second set--up one removes the piece of 
matter along the right beam. One observes the lifetime of the meter and thus
obtains $W\mathcal{W}$ information
for the object kaon as well. The coincidence counts of the
object and meter kaons show now no interference effects. They follow the 
non--oscillatory 
behaviour of Eqs.~(\ref{probS}) and (\ref{probL}).

Hence, we have constructed a quantum eraser allowing the
experimenter to erase or not the $W\mathcal{W}$ information by
placing or not the piece of matter along the right beam. The
first set--up shows the ``wave--like'' behaviour of the object
kaon, i.e., the two different components $K_S$ and $K_L$ are indistinguishable
because their marks are made inoperative by the strangeness measurement
on the meter kaon.
One gets interferences as in common double--slit experiments with indistinguishable paths.
The second set--up  clearly demonstrates the ``particle--like'' behaviour of the object
kaon: no interference is observed because the meter mark is operative and one
gains $W\mathcal{W}$ information on the right moving kaon. It mimics
double--slit set--ups with complete path information.

These experiments are analogous to the photon experiments
of Refs.~\cite{zeilinger95,walborn}, as discussed in detail in Ref.~\cite{BGH1}.
Note, however, that  in the kaon case the amplitudes are automatically marked and 
no double--slit is needed.

\subsection*{(b) Partially active eraser with \emph{active} measurements}
In this case one always inserts a piece of matter in the right hand beam
at a fixed time $\tau_r^0$, but the experiment is now also
designed to detect decays (any decay modes) occurring between the origin and this
piece of matter. In this way the right moving kaon ---the meter--- can make
the ``choice'' to show $W\mathcal{W}$ information by decaying before
$\tau_r^0$. If the meter kaon does indeed decay in free space, one measures its
lifetime \emph{actively} and obtains $W\mathcal{W}$ information. If no decay is seen,
the incoming kaon is projected into one of the two strangeness states at $\tau_r^0$ by  an
\emph{active} strangeness measurement.

With a single experimental set--up one observes the ``wave'' behaviour (interference)
for some events and the ``particle'' behaviour ($W\mathcal{W}$ information) for others.
The choice to obtain or not $W\mathcal{W}$ information
is naturally given by the instability of the kaons. However, the
experimenter can still choose when ---the time $\tau_r^0$---
he or she wants to learn something
about the strangeness of the meter system. Thus there is no control over the marking and
the erasure of individual kaon pairs, but a probabilistic prediction for an
ensemble of kaon pairs is known. We call this a partially active eraser experiment.

This experiment is analogous to the eraser experiment with entangled
photons of Ref.~\cite{scully00}. The role played by the beam--splitter
transmittivities in the photonic experiment is played by $\tau_r^0$ in the kaon case.

\subsection*{(c) Passive eraser with a \emph{passive} measurement}
In this experiment one is interested in the different decay modes of
the right moving meter kaon and one thus considers a \emph{passive}
measurement of strangeness or lifetime on it. Now one clearly has a
completely passive  erasing operation on the meter and the experimenter has
no control on the operativity of the lifetime mark. Only the object system is under some
kind of active control ---one still makes an \emph{active} strangeness
measurement and considers only kaons surviving up to this detector.
Remarkably, one finds
the same joint probabilities as in the previous cases (a) and (b).

\subsection*{(d) Passive eraser with \emph{passive} measurements}
This is the extreme case of a (passive) quantum eraser. The
experimenter has no control over any individual pair \emph{neither} on
which of the two complementary observables are going to be measured
\emph{nor} when they are measured. The experiment is also
totally symmetric and so it shows the full behaviour of the
maximally entangled state (\ref{time}). Remarkably, the
results of the observable probabilities are again in agreement with
all previous results. However, as nothing is actively measured or erased, it is hard to
speak about a quantum eraser. Clearly, there exists no analog for any other
spin--$1/2$ entangled system (except for the $B^0\bar B^0$
system\footnote{Due to the very short $B$--meson lifetime,
for the $B^0\bar B^0$ system only
\emph{passive} measurements are possible and clearly only a
passive eraser can be realized.}).

In particular, the joint probabilities for like-- and unlike--strangeness measurements
coincide with those in Eqs.~(\ref{lSprob}) and (\ref{uSprob}). They are measured by
counting and properly sorting the joint semileptonic decays occurring 
at different values of $\tau_l$ and at fixed $\tau^0_r$. 
The oscillatory behaviour on each one of these variables
is observed regardless $\tau_l$ is larger or smaller than $\tau^0_r$.
The same happens with the joint probabilities in Eqs.~(\ref{probS}) and (\ref{probL})
corresponding to strangeness--lifetime passive measurements. Their 
non--oscillatory behaviour is observed regardless of the time ordering
of $\tau_l$ and $\tau^0_r$. For kaons, the
delayed choice mode of a quantum eraser contains no additional mystery.
Time ordering is seen not to be the issue. It is the sorting of the various joint events,
irrespectively of any time consideration, which is crucial for quantum erasure.

\section{Conclusions}

Under the assumption of $CP$ conservation and the validity of the $\Delta S = \Delta Q$
rule, we have shown that kaons admit two alternative procedures to measure
their two complementary observables: strangeness and lifetime.  We
call these procedures \emph{active} and \emph{passive} measurements. The
first one can be seen as an analog to the usual von Neumann
projection, the second one is quite different and takes advantage of the information
spontaneously released by the neutral kaon decay modes.

We proposed four different experiments combining \emph{active} and
\emph{passive} measurement procedures and demonstrating the quantum erasure
principle for kaons. Remarkably, all four considered experiments lead to the same
observable probabilities and to the same physical results and ---more important for delayed
choice considerations--- this is true regardless of the temporal ordering of the
measurements. In our view, this illustrates the very nature of a quantum eraser
experiment: it essentially sorts different events, namely, strangeness--strangeness events
representing the ``wave'' property of the object or strangeness--lifetime events
representing the ``particle'' property of the object.

There are no experiments up to  date verifying the
proposed quantum marking and eraser ideas, except that the CPLEAR
collaboration \cite{CPLEAR} did  part of the job of our first
set--up (a) showing the entanglement of kaon pairs and measuring two points testing the
oscillatory behaviour of strangeness--strangeness joint detections.
We think that the proposed experiments are of interest because
they offer a new test of complementarity and shed new light on the very
concept of the quantum eraser.

\vspace{4mm}
\noindent {\bf Acknowledgements} \\
This work has been partly supported by EURIDICE HPRN-CT-2002-00311, BFM-2002-02588,
Austrian Science Foundation (FWF)-SFB 015 P06 and INFN.



\end{document}